\documentclass[prl,twocolumn,nopacs,nofootinbib]{revtex4}

\usepackage{bm}
\usepackage{bbm}
\usepackage{amsmath}
\usepackage{amssymb}
\usepackage{amsthm}
\usepackage{amsfonts}
\usepackage{mathtools}
\usepackage{latexsym}
\usepackage{relsize}
\usepackage[greek,english]{babel}
\usepackage{booktabs}
\usepackage[iso-8859-7]{inputenc}

\usepackage{array}
\newcolumntype{?}{!{\vrule width 2pt}}

\newcommand\postl{\stackrel{\mathclap{\normalfont\mbox{${\tiny \omega\rightarrow 0}$}}}{=}}
\newcommand\postlw{\stackrel{\mathclap{\normalfont\mbox{${\tiny \omega, K \rightarrow 0}$}}}{=}}
\newcommand\postt{\stackrel{\mathclap{\normalfont\mbox{${\tiny \Omega^A \rightarrow 0}$}}}{=}}

\newcommand{\lp}{\left(}
\newcommand{\rp}{\right)}
\newcommand{\lb}{\left[}
\newcommand{\rb}{\right]}

\newcommand{\lsim}   {\mathrel{\mathop{\kern 0pt \rlap
  {\raise.2ex\hbox{$<$}}}
  \lower.9ex\hbox{\kern-.190em $\sim$}}}
\newcommand{\gsim}   {\mathrel{\mathop{\kern 0pt \rlap
  {\raise.2ex\hbox{$>$}}}
  \lower.9ex\hbox{\kern-.190em $\sim$}}}
\newcommand{\bw}{\begin{widetext}\begin{equation}}
\newcommand{\ew}{\end{equation}\end{widetext}}
\newcommand{\be}{\begin{equation}}
\newcommand{\ee}{\end{equation}}
\newcommand{\ba}{\begin{eqnarray}}
\newcommand{\ea}{\end{eqnarray}}

\newcommand{\diff}{{\rm d}}
\newcommand{\Df}{{\rm D}}

\newcommand{\kz}{{\mathbbmtt{\kappa}}}
\newcommand{\oz}{{\mathbbmtt{\omega}}}
\newcommand{\ez}{\mathbbmtt{\theta}}
\newcommand{\ezv}{\mathbbmtt{(1/\theta)}}
\newcommand{\lzv}{\mathbbmtt{(1/\sigma)}}
\newcommand{\lz}{\sigma}
\newcommand{\fz}{h}

\newcommand{\al}{\alpha}
\newcommand{\bt}{\beta}

\newcommand{\J}{\mathbf{{\Omega}}}
\newcommand{\T}{\mathbf{T}}
\newcommand{\Sp}{\mathbf{S}}
\newcommand{\K}{\mathbf{K}}

\newcommand{\F}{\mathbf{F}}
\newcommand{\R}{\mathbf{R}}
\newcommand{\La}{\mathbf{\Lambda}}

\newcommand{\Km}{\mathbf{k}}
\newcommand{\Rm}{\mathbf{r}}
\newcommand{\Sm}{\mathbf{s}}
\newcommand{\Tm}{\mathbf{t}}

\newcommand{\as}{/\mathbf{a}/}

\newcommand{\fs}{/\mathbf{f}/}
\newcommand{\hs}{/\mathbf{h}/}
\newcommand{\gs}{/\mathbf{g}/}

\newcommand{\by}{\,\mathbf{b}}
\newcommand{\cy}{\,\mathbf{c}}
\newcommand{\fy}{\,\mathbf{f}}
\newcommand{\hy}{\,\mathbf{h}}
\newcommand{\gy}{\,\mathbf{g}}
\newcommand{\aq}{\backslash\mathbf{a}\backslash}

\newcommand{\fq}{\backslash\mathbf{f}\backslash}
\newcommand{\hq}{\backslash\mathbf{h}\backslash}
\newcommand{\gq}{\backslash\mathbf{g}\backslash}

\newcommand{\stara}{\prescript{\star}{}}

\begin{document}

\title{On the equivalence principle as a breaking of the dilatation symmetry}

\author{Tomi Sebastian Koivisto}
\email{tomik@astro.uio.no}
\affiliation{Nordita, KTH Royal Institute of Technology and Stockholm University,
Roslagstullsbacken 23, 10691 Stockholm, Sweden}
\affiliation{Helsinki Institute of Physics, P.O. Box 64, FIN-00014 Helsinki, Finland} 
\affiliation{Department of Physical Sciences, Helsinki University, P.O. Box 64, FIN-00014 Helsinki, Finland}

\preprint{NORDITA-2016-37}
\date{\today}

\begin{abstract}

The so called Telelateral Conformal Gravity (TCG) is a first order gauge theory describing inertia as internal, scale invariant geometry. 
The Conformal Affine Theory (CAT) unifies other interactions into TCG in a minimalistic scheme via an exact holographic gauge-gravity duality. 
In the three parts of this Letter, the currently topical bimetric gravity models are obtained as non-inertial gauges in the CAT,
the geometrisation of electromagnetism by Hermann Weyl is brought to a completion, and it is sketched
how the rest of the Standard Model could result from the assumption of a light cone, respectively.
The radically intuitive CAT suggests the possibility of further transgressive unification of sciences.

\end{abstract}

\maketitle

It can be argued that the operational significance of the equivalence principle is that an observer may chart the space with Cartesian coordinates. 

In the conventional formulation originally due to Einstein, it follows from the identification of the gravitational and inertial mass, that the curvature of spacetime disappears in a ''free fall'',  that can be recovered by a coordinate transformation that reduces the metric $g_{\mu\nu}$ to\footnote{We use the signature $\eta_{\mu\nu}=(-1,1,1,1)$,
and the following notations for the indices in their respective spaces: \\
\begin{tabular}[b]{l l l}
\hspace{0cm}$a,b = 0,...,3$ &  group &   $\eta_{ab}$ \\
\hspace{0cm}$A,B =1,...,15$ & FUNDAMENTAL  &  $ h_{AB} = f^{M}_{\phantom{M}AN} f^{N}_{\phantom{N}BM}$ \\
\hspace{0cm}$\al,\bt = 0,...,3 $ & \greektext{qwroqr\'onou} &  $f_{\al\bt}$\,, $g_{\mu\nu}$\,, $h_{\mu\al}$, $G_{\mu\al}$ \\
\hspace{0cm}$\mathbf{a},\mathbf{b} = 0,...,3 $ & $\mathbf{meta}$ & $\hat{\mathbf{\delta}}_{(\mathbf{a}\mathbf{b})}\,.$
\end{tabular}
\vspace{0cm}
}
$\eta_{\mu\nu}$. 

The modern gauge perspective however allows equivalent but alternative interpretations. 

{\it Teleparallel} gravity arises naturally as a gauge theory of the translation, where the spin connection sources inertia  \cite{Aldrovandi:2013wha}. In a frame where the metric becomes $\eta_{\mu\nu}$, the gravitational force is exactly canceled by the inertia. 

In the {\it telelateral}\,\footnote{A.k.a the ''symmetric teleparallel gravity'' \cite{Nester:1998mp,Adak:2008gd}, since the gauge is torsionless but also teleparallel, i.e. curvature-free. Canonical projection to spacetime would be the Palatini variation \cite{Amendola:2010bk,Koivisto:2011vq} with the $g_{\mu\nu}$ set to $\eta_{\mu\nu}$. As the gravitational interaction is internalised, the connection does not preserve the flat metric.} gauge,
 gravity operates in the non-metricity \cite{Mol:2014ooa} sector of the larger symmetry, affinity \cite{Hehl:1994ue}. The conventional picture is reversed in the way that a physical spacetime is {\it a priori} geometrically trivial, but however ''non-inertially'' connected due to gravity, an aspect of internal geometry.

It was recently pointed out that the Weyl curvature action becomes first order in the latter gauge \cite{Frank}. Telelateral Conformal Gravity (TCG) could thus be a unitary and renormalisable theory. The metric $g_{\mu\nu}$ arising as an integral of the connection $\La$ (vs being a potential for a derivative force), the nonlocal \cite{Biswas:2011ar} freezing of the bulk dynamics \cite{Marolf:2014yga} and the thermodynamic nature of gravity \cite{Jacobson:1995ab} become more apparent in the telelateral gauge. 

The focus of this letter, however, is to describe how the Conformally Affine Theory (CAT) corroborates the rationale of TCG \cite{Frank} by incorporating the rest of the natural symmetries within it.  
We are lead to contemplate Special Relativity as a definition of time, in the sense of a causal structure expressed in the algebra of the six Lorentz generators $\J_{ab}=-\J_{ba}$,
\be \label{cr1}
[\J_{ab},\J_{cd}]   =   2\lp \eta_{d[a}\J_{b]c} -  \eta_{c[a}\J_{b]d}\rp\,,
\ee
and General Relativity in essence as an invocation of space, by the translation generators $\T_a$,  
\be \label{cr2}
[\T_a,\J_{bc}]  =  2\eta_{a[b}\T_{c]}\,, \quad [\K,\T_a] = -\T_a\,,
\ee 
with the dilatation $\K$ present in the conformal gauge theory. In the CAT, time and space together with the Standard Model of Particle Physics (SM) are implied by the postulate of a null cone. 
The claim is that it is not a coincidence that the background symmetry of the SM (at high energies) is the same as the gauge symmetry of the TCG, but a manifestation of a profound holography, a circular reasoning built into the equations. 

To be precise, the null cone is determined by the $\K$ introduced above, and we will gauge\footnote{On the gauging of the special orthogonal group (4,2), see e.g. classic papers \cite{MacDowell:1977jt,Kaku:1977pa}, systematic presentations \cite{Wheeler:1997pd,Trujillo:2013saa} and current developments \cite{Attard:2015jfr,Zlosnik:2016fit}.} also the special conformal transformation $\Sp_a$. The origin of spacetime and its material contents is then the statement $\kappa^2=0$, c.f. (\ref{null}), $\kappa \K$ being the dilation in the fundamental connection $\La$.

We point attention to the similarity between the commutation rules for the generators $\T_a$ in (\ref{cr2}) and those for $\Sp_a$:
\be \label{cr3}
[\Sp_a,\J_{bc}]   =    2\eta_{a[b}\Sp_{c]}\,, \quad [\K,\Sp_a] = \Sp_a\,.
\ee
It is understood that the special conformal transformation does not reside in the non-metricity sector of the affinity together 
with the dilatation, but in an inhomogeneous sector as an oppositely-dilatating dual of the translation. The commutation of the two, 
\be \label{cr4}
 [\T_a,\Sp_b]=2\lp\eta_{ab}\K-\J_{ab}\rp\,,
\ee
produces the generators of the homogeneous sector. 

The fundamental connection $\La$, generated linearly by the $\mathbf{G}_A$ defined above, then entails the associated four gauge potentials $\Lambda^A=\{\kz,\oz^{ab},\lz^a,\ez^a\}$ as:
\be 
\La = \Lambda^A {\mathbf{G}_A} = \kz \K + \oz^{ab}\mathbf{\Omega}_{ab} + \lz^a \Sp_a + \ez^a\T_a \,.
\ee
The relations (\ref{cr1},\ref{cr2},\ref{cr3},\ref{cr4}) specify completely the non-vanishing structure constants $f^{A}_{\phantom{A}BC}$ of the Lie algebra,
\be
[{\mathbf{G}_A},{\mathbf{G}_B}] = f^{C}_{\phantom{C}AB}{\mathbf{G}_C}\,,
\ee
and we can determine the field strengths by the Maurer-Cartan (de)construction (second application of which would give the Bianchi identities) with
the $\Lambda$-covariant (exterior) derivative $\Df$:
\ba
K & = &  \Df \kz = \diff\kz + \lz_c\wedge\ez^c\,, \\
\Omega^{ab} & = &  \Df \oz^{ab} = \diff\oz^{ab}+\oz^{a}_{\phantom{a}c}\wedge \oz^{cb} + \lz^{[a}\wedge \ez^{b]}\,, \\
S^a & = &  \Df \lz^{a} = \diff\lz^a + \oz^{a}_{\phantom{a}c}\wedge\lz^c + \kz\wedge\lz^a\,,  \label{ds}\\  
T^a & = &  \Df \ez^{a} = \diff\ez^a + \oz^{a}_{\phantom{a}c}\wedge\ez^c - \kz\wedge\ez^a \label{dt}\,. 
\ea
To form an action for these two-forms we need to project them into a spacetime.

\section{Space}


Now we have at hand two sets of translation generators, and can consider setting up spacetimes from the 
various distinct metrics that emerge, 
\be \label{triad}
f_{\al\bt} =  \eta_{ab}\lz^a_\al\lz^b_\bt\,, \quad 
g_{\mu\nu}  =  \eta_{ab}\ez^a_\mu\ez^b_\nu\,, \quad  
h_{\mu\bt}  =  \eta_{ab}\ez^a_{(\mu}\lz^b_{\bt)}\,,
\ee
and from the various volume elements $\diff^4 V_{\backslash\mathbf{a}\backslash\backslash\mathbf{b}\backslash}{}$ that can be formed by contracting the supposed spacetime coordinate differentials $\lp \diff x^\mu  \diff x^\nu  \diff x^\rho  \diff x^\sigma\rp$ with the $\epsilon_{abcd}\ez^a_\mu\ez^b_\nu\ez^c_\rho\ez^d_\sigma$ and its generalisations as follows: 
\vspace{0.25cm}\\
\begin{tabular}{ll l l}
\hspace{0.25cm}$\epsilon_{abcd}\lz^a\lz^b\lz^c\lz^d \rightarrow\diff^4V_{\fq\fq}{}$ &   : special  \\
\hspace{0.25cm}$\epsilon_{abcd}\lz^a\lz^b\lz^c\ez^d \rightarrow\diff^4V_{\fq\hq}{}$ &  : odd special      \\
\hspace{0.25cm}$\epsilon_{abcd}\lz^a\lz^b\ez^c\ez^d \rightarrow\diff^4V_{\fq\gq}{}=\diff^4V_{\hq\hq}{}$ &   : partial  \\
\hspace{0.25cm}$\epsilon_{abcd}\lz^a\ez^b\ez^c\ez^d \rightarrow\diff^4V_{\gq\hq}{}$ &  : odd spatial   \\
\hspace{0.25cm}$\epsilon_{abcd}\ez^a\ez^b\ez^c\ez^d \rightarrow\diff^4V_{\gq\gq}{}$ &  : spatial\,. 
\end{tabular}
\vspace{0.25cm}\\
We can introduce triads of tensor-valued two-forms consisting of the various projections of the curvatures, the simplest one being 
\be \label{km}
\Km  =  \prescript{\as}{}k_{\al\bt} \lp \diff^2 V^{\alpha\beta}_{\backslash\mathrm{a}\backslash{}}\rp\,, \quad  \prescript{\as}{}k_{\al\bt} = K_{\al\bt}\,.
\ee
The planting the Lorentz field strength into different area elements gives rise to a matrix $\Rm^{\mathbf{a}\mathbf{b}}$ of similar triads, and the matrix can be projected to form tensor-valued four-forms. We obtain three distinct such objects, which are generalisations of the familiar Riemannian curvature four-form:
\begin{subequations}
 \label{curvatures}
\ba
\Rm_{\mathbf{f}} & = &  h_{\gamma\delta}h_{\kappa\lambda}\lzv^\delta_b\lzv^\lambda_c\prescript{\as}{}\Omega^{bc}_{\al\bt}\diff^4 V^{\al\bt{\gamma}{\kappa}}_{\aq\fq} \,, \\
\Rm_{\mathbf{g}}  & = &  g_{\mu\rho}g_{\nu\sigma}\ezv^\rho_b\ezv^\sigma_c\prescript{\as}{}\Omega^{bc}_{\al\bt}\diff^4 V^{\al\bt\mu\nu}_{\aq\gq}\,, \\
\Rm_{\mathbf{h}}  & = &  g_{\mu\rho}f_{{\gamma}\delta}\ezv^\rho_b\lzv^\delta_c\prescript{\as}{}\Omega^{bc}_{\al\bt}\diff^4 V^{\al\bt\mu{\gamma}}_{\aq\hq}\,.
\ea
\end{subequations}
For now, we assume that the torsional field strengths step only into their own territory. 
The only non-vanishing projections of (\ref{ds}) and (\ref{dt}) are then 
\label{torsions}
\be \label{torsions}
\Sm^{\mathbf{a}}  =  \lz^a_\gamma S^{\gamma}_{\phantom{\mu}\al\bt}\diff^2 V^{\al\bt}_{\fq}\,, \quad
\Tm^{\mathbf{a}}  =  \ez^a_\mu T^{\mu}_{\phantom{\mu}\rho\sigma}\diff^2 V^{\rho\sigma}_{\gq}\,, 
\ee
where the components are 
\ba
S^{\gamma}_{\phantom{\mu}\al\bt} & = &   2\lzv^\gamma_a\lp\lz^a_{[\bt,\al]} + \oz^a_{\phantom{a}c[\al} \lz^c_{\bt]} \rp - 2\delta^{\mu}_{[\al}\kz_{\bt]}\,,  \label{S} \\
T^{\mu}_{\phantom{\mu}\rho\sigma}  &= &   2\ezv^\mu_a\lp\ez^a_{[\sigma,\rho]} + \oz^a_{\phantom{a}c[\rho} \ez^c_{\sigma]} \rp + 2\delta^{\mu}_{[\rho}\kz_{\sigma]}\,. \label{T} 
\ea 
To be specify the summation over the bold indices, we parameterise an underlying metric,
\be
\hat{\mathbf{\delta}}_{\mathbf{a}\mathbf{b}} = 
\left( \begin{array}{ccc} \label{delta}
a & d & c \\
d & b & e \\
c & e & c \end{array} \right)\,,
\ee
with five distinct components.

The four curvature terms in (\ref{km},\ref{curvatures},\ref{torsions}) can now be combined into the general quadratic Lagrangian four-form in a diagonal basis\footnote{The objects (\ref{km},\ref{curvatures},\ref{torsions}) are multiplied with the wedge product, so that the 4-form components of (\ref{quadraction}) can be written 
\ba
\prescript{/\mathbf{a}//\mathbf{b}/}{}{}\mathrm{L}  & = &
 \prescript{/\mathbf{a}/}{}\Km\wedge  \prescript{/\mathbf{b}/}{}\Km^\star 
+ \alpha_R \hat{\mathbf{\delta}}_{\mathbf{c}\mathbf{e}}\hat{\mathbf{\delta}}_{\mathbf{d}\mathbf{f}}\prescript{/\mathbf{a}/}{}\Rm^{\mathbf{cd}}\wedge \prescript{/\mathbf{b}/}{}{\Rm^{\mathbf{ef}}}^\star \nonumber
 \\ & + & 
 \mathbf{\delta}_{\mathbf{c}\mathbf{d}}\lp m_{S}^2\prescript{/\mathbf{a}/}{}\Sm^{\mathbf{c}}  \prescript{/\mathbf{b}/}{}\Sm^{\mathbf{d}\star}
+ m_{T}^2 \prescript{/\mathbf{a}/}{}\Tm^{\mathbf{c}}\wedge  \prescript{/\mathbf{b}/}{}\Tm^{\mathbf{d}\star}\rp \,. \nonumber
\ea},
\be \label{quadraction}
\mathrm{L} =  \prescript{/\mathbf{a}//\mathbf{b}/}{}{}\mathrm{L}\diff^4V_{\backslash\mathrm{a}\backslash{}\backslash\mathrm{b}\backslash}  =  \Km^2 + \alpha_R\Rm^2 + m_{S}^2 \Sm^2 + m_{T}^2 \Tm^2\,.
 \ee
Expanding in terms of the curvatures (\ref{km},\ref{curvatures}), we have
\ba
\mathrm{L} & = &   m_S^2 \Sm^2 + \alpha_f \lp\Rm^{\fy}\rp^2 + aK^2\diff^4 V_{\fq\fq} \nonumber \\
 & + & m_T^2 \Tm^2 +
            \alpha_g\lp\Rm^{\gy}\rp^2 +  bK^2\diff^4 V_{\gq\gq} \nonumber \\
            & + & \alpha_h\lp \Rm^{\hy}\rp^2 + 3c K^2\diff^4 V_{\hq\hq} \nonumber \\
            & + &  2 K^2\lp d\diff^4 V_{\fq\gq}  + e\diff^4 V_{\fq\hq}\rp\,,  \label{quadraction2}
\ea
where 
\ba 
\alpha_f & = & \alpha_R \lp a + b + c\rp^2\,, \quad \alpha_g = \alpha_R\lp b + 3c + e + d \rp^2\,, \nonumber \\ 
\alpha_h & = & 2\alpha_R   \lp a + b + c\rp \lp b + 3c + e + d \rp = 2\sqrt{\alpha_f\alpha_g}\,. \label{alphas}
\ea
We can separate the post-Lorentzian part of the spacetime-projected gauge curvatures in (\ref{curvatures}) as
\begin{subequations}
\label{curvatures2}
\ba   
\prescript{\fs}{}\Omega_{\al\bt\gamma\delta} & = & \mathcal{R}^{\mathbf{f}}_{\al\bt\gamma\delta} + \frac{1}{2}\lp f_{\al[\gamma}h_{\delta]\bt} - f_{\beta[\gamma}h_{\delta]\alpha}\rp\,, \\\
\prescript{\gs}{}\Omega_{\rho\sigma\mu\nu} & = & \mathcal{R}^{\mathbf{g}}_{\rho\sigma\mu\nu} - \frac{1}{2}\lp g_{\rho[\mu}h_{\nu]\sigma} - g_{\sigma[\mu}h_{\nu]\rho}\rp\,, \\
\prescript{\hs}{}\Omega_{\mu\al\nu\bt} & = & \mathcal{R}^{\mathbf{h}}_{\mu\alpha\nu\beta} - \frac{1}{2}\lp h_{\mu[\nu}h_{\bt]\al} - h_{\al[\nu}h_{\bt]\mu}\rp\,, 
\ea 
\end{subequations}
where $ \mathcal{R}^{\mathbf{a}}_{\al\bt\mu\nu}$ is the part of the curvature that vanishes identically in the gauge $\omega=\partial\omega=0$. 
We note that the algebraic frame couplings can be rewritten in terms of the symmetric polynomials of the metric determinants:
denoting $\det{\lp h_{\mu\nu}\rp}=[h]$, $\det{\lp h_{\mu\al}h^{\al}_{\phantom{\al}\nu}\rp} = [h^2]$,
$ \det{\lp h_{\mu\al}h^{\al\bt}h_{\bt\nu}\rp}= [h^3]$,
the post-Lorentzian contributions from (\ref{curvatures2}) in (\ref{quadraction2}) read
\begin{subequations} \label{hrterms}
\ba
\lp\Rm^{\fy}\rp^2 \quad & \postl & \quad 6\lb h\rb \diff^4 V_{\gq\gq} \equiv \mathrm{r}_1(h)V_{\gq\gq}\,, \\
 \lp\Rm^{\gy}\rp^2 \quad & \postl & \quad  2\lp\lb h\rb^2-\lb h^2\rb\rp \diff^4 V_{\gq\gq} \nonumber \\ & \equiv & \mathrm{r}_2(h)V_{\gq\gq}\,, \\
 \lp\Rm^{\hy}\rp^2  \quad & \postl & \quad \lp\lb h\rb^3-3\lb h\rb \lb h^2\rb+2\lb h^3\rb\rp \diff^4 V_{\gq\gq} \nonumber \\ & \equiv & \mathrm{r}_3(h)V_{\gq\gq}\,,
\ea
\end{subequations}
where the $\mathrm{r}_1(h)$, $\mathrm{r}_2(h)$ and $\mathrm{r}_3(h)$ are the symmetric polynomials.
These represent the three non-trivial dRGT/Hassan-Rosen terms that appear in the ghost-free massive and bimetric models of gravity \cite{deRham:2014zqa,Schmidt-May:2015vnx}, with the canonical coefficients $\beta_1$, $\beta_2$, $\beta_3$ respective to the Einstein-Hilbert term when $\alpha_R = \frac{1}{8}(m_g/m_T)^2$, where $m_g$ is the graviton mass parameter. 
The cosmological constants, $\beta_0$ and $\beta_5$ would be given by a non-vanishing expectation value $\kappa^2$, but let us 
freeze the Weyl vector to the classical value $\kappa_\mu=0$ and express the torsion invariants in terms of the equivalent (up to a total integral) metric Ricci scalars, to finally obtain:
\ba \label{bimetric}
\mathrm{I} \quad\quad  & \postlw & \quad\quad -\int  \diff^4 x \sqrt{-f} \frac{m_S^2}{2}\mathcal{R}^{\mathbf{f}} \nonumber \\ & - &  \int  \diff^4 x\sqrt{-g}\lp \frac{m_T^2}{2}\mathcal{R}^{\mathbf{g}}-\alpha_R \sum_{n=1}^3\beta_n \mathrm{r}_n(h)\rp\,,
\ea 
where $\beta_1=6\alpha_f$, $\beta_2=2\alpha_g$ and $\beta_3=4\sqrt{\alpha_f\alpha_g}$, with $\alpha_f$ and $\alpha_g$ given by (\ref{alphas}). These couplings vanish in the teleparallel gauge, wherein the gauge curvature of the Lorentz transformation is set to zero, $\Omega^{ab}=0$. Here they appear as post-Lorentzian curvature that is due to the coupling of two distinct translational gauge fields, the spatial and the special, via the rotational curvature. The interaction can thus be seen as an inertial effect\footnote{Starting from the bigravity side, conformality has been looked for \cite{Deser:2012qg,Hassan:2015tba}. It has been also pointed out that bimetric \cite{Akrami:2014lja} cosmology (\ref{bimetric}) is (Q)CD \cite{Isham:1971gm} in the sky \cite{Koivisto:2015qum}.}.

\section{Time}

Let us now consider the quadratic theory (\ref{quadraction}) instead in the teleparallel gauge where the total curvature vanishes\footnote{The $\omega\rightarrow 0$ gauge could perhaps be called the Weitzenb\"ock frame to distinguish from the teleparallel frame $\Omega^A\rightarrow 0$. They coincide if the gauge group is the Poincar\'e.}.  

In the two displacement sectors of the geometry, $T$ and $S$, we can define the torsion $T^\rho_{\phantom{\rho}\mu\nu} = -T^\rho_{\phantom{\rho}\nu\mu}$, its vector component $T^{V}_\mu=T^\rho_{\phantom{\rho}\mu\rho} =-T^\rho_{\phantom{\rho}\rho\mu}$, axial component $T^A_\mu$, superpotential $\tilde{T}^\rho_{\phantom{\rho}\mu\nu} = -\tilde{T}^\rho_{\phantom{\rho}\nu\mu}$ and contortion $\Gamma^T_{\rho\mu\nu} =-\Gamma^T_{\nu\mu\rho}$, respectively, as follows:
\begin{subequations}
\label{torsiontensors}
\ba
T_{{\rho}\mu\nu} & = &  \mathcal{T}_{{\rho}\mu\nu} + 2qg_{\rho [\mu}\kappa_{\nu]}\,, \\
T^V_\mu & = & \mathcal{T}^V_\mu - 3q\kappa_\mu\,, \quad T^A_\mu  =  \frac{1}{6}\epsilon_{\mu\nu\rho\sigma}\mathcal{T}^{\nu\rho\sigma}\,, \\
\Gamma^T_{\rho\mu\nu} & = & \frac{1}{2}\lp \mathcal{T}_{\rho\mu\nu}+\mathcal{T}_{\nu\rho\mu} - \mathcal{T}_{\mu\nu\rho}\rp  - 2qg_{\mu [\nu}\kappa_{\rho]}\,, \\
\tilde{T}_{\rho\mu\nu} & = & \Gamma^T_{\mu\rho\nu} + 2g_{\rho[\mu}T^v_{\nu]} = \tilde{\mathcal{T}}_{\rho\mu\nu} - 4q g_{\rho [\mu}\kappa_{\nu]}\,. 
\ea
\end{subequations}
The curly symbols stand for the tensors in the Riemann-Cartan geometry, and we have thus explicitly separated the post-Poincar\'e contributions. For the $T$-sector $q=1$, and the $S$-sector is obtained from the above by $(T,\mathcal{T},q=1) \rightarrow (S,\mathcal{S},q=-1)$.
With the superpotential prescription of the Hodge dual for soldered bundles \cite{Lucas:2008gs}, the quadratic combinations appearing in the Lagrangian (\ref{quadraction}) correspond to
\be 
\frac{1}{2}\lp \tilde{T}T \rp \equiv \frac{1}{2}\tilde{T}_{\rho\mu\nu}T^{\rho\mu\nu}  =  \frac{1}{2}\tilde{\mathcal{T}}\mathcal{T} + 4q\mathcal{T}^V_\mu \kappa^\mu - 6 \kappa^2\,,
\ee
and so the (suitably normalised) action (\ref{quadraction}) can now be written as 
\ba
\mathrm{I} \quad &  \postt & \quad  \int\diff^4  V_{\hq\hq}  \lp F_{\mu\al} + \Sigma_{\mu\al}\rp   \lp F^{\mu\al} + \Sigma^{\mu\al}\rp    \nonumber \\
& + &  \frac{m_S^2}{2}\int\diff^4 V_{\fq\fq}   \lp {\mathcal{S}}\tilde{\mathcal{S}}   - 4\lp 2\mathcal{S}^V+3\kappa\rp\cdot\kappa\rp  \nonumber \\
& + &   \frac{m_T^2}{2}\int\diff^4 V_{\gq\gq}\lp {\mathcal{T}}\tilde{\mathcal{T}}   + 4\lp 2\mathcal{T}^V-3\kappa\rp\cdot\kappa\rp \label{taction}  \,.
\ea
We have planted the dilatation curvature into the neutral space for simplicity, and defined the field strength $F_{\mu\al}$ and the antisymmetric tensor $\Sigma_{\mu\al}$ as
\be
F_{\mu\al} = \partial_\mu \kappa_\al - \partial_\al\kappa_\mu\,, \quad \Sigma_{\mu\bt}  =  \eta_{ab}\ez^a_{[\mu}\lz^b_{\bt]}\,.
\ee
For the present purposes, we choose the Deser - van Nieuwenhuizen gauge condition $\Sigma_{\mu\beta}=0$.

The following gauge choice is now of a fundamental relevance: $\kappa^2=0$. That is, both of the
metrics are aligned with the gauge field of the dilatation,
\be \label{null}
f_{\alpha\beta}\kappa^{\alpha}\kappa^{\beta} = 0\,, \quad g_{\mu\nu}\kappa^{\mu}\kappa^{\nu} = 0\,.
\ee
This means that photons define the light cones for both of the metrics. To elaborate on the interpretation, we have not broken the 
gauge invariance of the theory, but rather by setting (\ref{null}) consider it in an intuitive gauge wherein the conventional definitions
of time are meaningful. In this sense, $\kappa^2=0$ is the equivalence principle in CAT. 

In the symmetric theory, $m_S=m_T$, the doublet of frame fields enjoys an exact exchange symmetry, except that, remarkably,
the couplings of the traces of the torsions cancel each other. Up to anomalous, symmetry breaking fluctuations then, we have $\ez^a=\lz^a$ and
the action (\ref{taction}) becomes (teleparallel) General Relativity coupled to Maxwell electrodynamics, if $m_S=m_T$ is adjusted to the suitable
ratio of the fine structure constant and the Planck mass.

\section{Matter}

The question arises what happens to the symmetry when couplings are taken into account. Let us, to be concrete, consider the gravitational coupling
of a spinor field \cite{Mosna:2003rx} $\Psi(x^a)$ with the mass $m$. We can now consider two projections of this field, $\Psi(x^\alpha)$ and $\Phi(x^{\mu})$,
in the two spacetimes spanned by $\lz^a$ and $\ez^a$, respectively.  The Dirac equations are
\begin{subequations}
\label{spinors}
\ba 
 i\lzv^\al_a\gamma^a \lp \partial_\al + \frac{1}{2}S^V_{\al} -\frac{3i}{2}S^A_\al \gamma^5\rp\Psi & = & m\Psi \,, \\
  i\ezv^\mu_a\gamma^a \lp \partial_\mu + \frac{1}{2}T^V_{\mu} -\frac{3i}{2}T^A_\mu \gamma^5\rp\Phi& = & m\Phi\,, 
\ea  
\end{subequations}
where $S^A_\al$ and $T^A_\mu$ are the axial components of the torsions. We can separate the post-Weitzenb\"ockian piece 
of the covariant derivative in the spinorial representation as follows (with the exhange notation from (\ref{torsiontensors}) restored):
\be \label{fock}
D_\mu = \partial_\mu + \frac{1}{2}\mathcal{T}^V_{\mu} -\frac{3i}{2}\mathcal{T}^A_\mu  + \frac{3}{2}q\kappa_\mu = \nabla_\mu + \frac{3}{2}q\kappa_\mu\,.
\ee
Note that in the Riemannian gauge, the connection $\nabla_\mu$ is just the Christoffel symbol.
We can thus identify $\Phi=\Psi^-$ as the antiparticle of $\Psi=\Psi^+$, since, in the symmetric case and in units where the
electric charge is $e=3/2$ we have
\be
i\gamma^\mu\lp \nabla_\mu \pm ie\kappa_\mu \rp\Psi^{\pm} = m\Psi^{\pm}\,.
\ee
Taking into account that the axial term in the Fock-Ivanenko derivative  (\ref{fock}) is complex, the theory is completely Charge-Parity-Time symmetric. Though we started with spinors in (\ref{spinors}), it is straightforward to reach the same conclusion for any other representation of charged fields.

Leptons can arise due to breaking of the residual symmetry in aligning the dilatation vector $\kappa_\mu$ in the two frames after setting (\ref{null}). The stability group of a null vector is
SO(3), out of which we can now double-solder the rest of the electroweak symmetry SU(2)$\times$U(1).

Mesons and baryons reflect fluctuations that violate Lorentz symmetry. In CAT, the immaculate sphericality
SO(4) within the Euclidean conformal group can be bi-projected into SU(3). Because in order to form spacetime representations
we have to pick three ''distance elements'' each from either the $S$-sector or the $T$-sector, quarks come with the electric charge of $\pm e/3$.

A more final formulation of the theory may be specified by the canonical Lagrangian four-form: 
\ba 
\frac{1}{2}\text{Tr}\lp\F^2\rp & = & 
\frac{1}{2}\text{Tr}\lp{\mathbf{G}_A} {\mathbf{G}_B}\rp F^A \wedge \prescript{\star}{}F^B = 
\frac{1}{2}h_{AB}F^A \wedge \stara F^{ B} \nonumber \\
& = &  K\wedge\stara K   - 2\Omega^{[ab]}\wedge \stara  \Omega_{[ab]} -  T^a\wedge \stara S_a\,. \label{canonical}
\ea
To contract indices of this four-form, one could introduce the following metric:
\ba
G_{\mu\al} & = & 
-\frac{1}{4}h_{AB}\lp \Lambda^A\rp^a_{\mu} \lp \Lambda^B\rp^{b}_{\al}\eta_{ab} \nonumber \\
& = &  - \frac{1}{2}\kappa_\mu\kappa_\al + 2\eta_{a[c}\eta_{d]b}\oz^{ab}_\mu  \oz^{cd}_\al  + h_{\mu\al}\,. \label{metric}
\ea 
Unlike with conventional solderings, the external theory then retains gauge (and not only diffeomorphism) covariance and the action its invariance under the change of the internal group basis. In the teleparallel gauge and with the identification of the photon (\ref{null}), the inverse metric that enters the action becomes 
\be
G^{\mu\al} \quad \postl \quad h^{\mu\al} + \frac{1}{2}\kappa^\mu\kappa^\al\,.
\ee
The first term is simply the conformally invariant, i.e. electrically neutral, ''mixed metric'', whose Einstein-Hilbert term indeed features in the canonical 
Lagrangian (\ref{canonical}). The photon term contracts to zero with any metric, but is not totally innocuous. -This, and the details of our scheme in 
the SU(2) and SU(3) sectors, will be discussed elsewhere.

CAT invites us to accept a more elastic concept of reality.
We may now understand why the same particle is responsible for our abstract concept of time and our visual experience, and the 
connection between the Wigner's problem and the fine-tuning required for the cosmological initial conditions. 
Spacetime singularities and horizons can be taken seriously as features of solutions that are extendable beyond the breaking points of the condition (\ref{null}),
''outside'' spacetime. We believe the implied conflict with unitarity is removed together with the observer-independent extrapolations of the notions of time and information. 


The nature of physics as the ultimate science of the observer is disclosing.



\bibliography{CLetter}

\end{document}